\documentclass[reprint, amsmath,amssymb, aps, showpacs, superscriptaddress, pra]{revtex4-1}

\usepackage{graphicx}
\usepackage{mathtools}
\usepackage{tensor}
\usepackage{verbatim}
\usepackage[breaklinks=true,colorlinks=true,linkcolor=blue,urlcolor=blue,citecolor=blue]{hyperref}
\usepackage{xcolor}
\usepackage{mdframed}

\input{Qcircuit}

\def\expect#1{\left\langle #1 \right\rangle}
\def\bra#1{{\langle #1 |}}
\def\ket#1{{| #1 \rangle}}

\newcommand{\C}{\textsc{C}}

\begin{document}
\preprint{}

\title{Ancilla-driven quantum computation for qudits and continuous variables}
\author{Timothy Proctor}
\thanks{Now at Sandia National Laboratories, Livermore, CA 94550, USA}
\affiliation{School of Physics and Astronomy, E C Stoner Building, University of Leeds, Leeds, LS2 9JT, UK}
\affiliation{Berkeley Quantum Information and Computation Center, Department of Chemistry, University of California, Berkeley, CA 94720, USA}
\author{Melissa Giulian}

\author{Natalia Korolkova}
\affiliation{School of Physics and Astronomy, University of St. Andrews,
North Haugh, St.~Andrews, Fife, KY16 9SS, Scotland}
\author{Erika Andersson}
\affiliation{Institute of Photonics and Quantum Sciences, School of Engineering \& Physical Sciences, Heriot-Watt University, Edinburgh, EH14 4AS, United Kingdom}
\date{\today}
\author{Viv Kendon}
 \email{viv.kendon@durham.ac.uk}
\affiliation{Department of Physics, Durham University, Durham, DH1 3LE, UK}
\date{\today}

\begin{abstract}
Although qubits are the leading candidate for the basic elements in a quantum computer, there are also a range of reasons to consider using higher dimensional qudits or quantum continuous variables (QCVs). In this paper we use a general `quantum variable' formalism to propose a method of quantum computation in which ancillas are used to mediate gates on a well-isolated `quantum memory' register and which may be applied to the setting of qubits, qudits (for $d>2$) or QCVs. More specifically, we present a model in which universal quantum computation may be implemented on a register using \emph{only}: repeated applications of a single fixed two-body ancilla-register interaction gate, ancillas prepared in a single state, and local measurements of these ancillas. In order to maintain determinism in the computation, adaptive measurements via a classical-feedforward of measurement outcomes are used, with the method similar to that in measurement-based quantum computation (MBQC). We show that our model has the same hybrid quantum-classical processing advantages as MBQC, including the power to implement any Clifford circuit in essentially one layer of quantum computation. In some physical settings, high-quality measurements of the ancillas may be highly challenging or not possible, and hence we also present a globally unitary model which replaces the need for measurements of the ancillas with the requirement for ancillas to be prepared in states from a fixed orthonormal basis. Finally, we discuss settings in which these models may be of practical interest.
\end{abstract}


\maketitle

\section{Introduction}
There are some compelling reasons to consider implementing a quantum computer with higher dimensional \emph{qudits} ($d$-level systems, $d>2$) or systems with a continuous degree of freedom -- \emph{quantum continuous variables} (QCVs). Particularly interesting recent results show that fault-tolerance thresholds for $d$-dimensional qudits are improved by increasing $d$ \cite{watson2015qudit,campbell2014enhanced,anwar2014fast,campbell2012magic,andrist2015error,duclos2013kitaev} and it is also known that increasing the dimension of the qudits can improve the robustness of some algorithms \cite{parasa2011quantum,zilic2007scaling,parasa2012quantum} and provide a logarithmic decrease in the number of subsystems required for a computation \cite{Muthukrishnan2000Multivalued,stroud2002quantum}. Furthermore, high-quality quantum controls over $d>2$ qudits have now been experimentally demonstrated in a variety of physical settings \cite{neeley2009emulation,smith2013quantum,anderson2014accurate,bent2015experimental,walborn2006quantum,lima2011experimental,rossi2009multipath,dada2011experimental}, providing additional motivation for research into qudit-based quantum computers. Turning to QCVs, in the optical settings these are some of the easiest quantum systems to entangle and manipulate, as demonstrated by a range of impressive experiments \cite{ukai2011demonstration,su2013gate,jensen2011quantum,yokoyama2013ultra}, including the creation of entangled states of 10,000 individually addressable QCVs \cite{yokoyama2013ultra}. Moreover, although QCVs may seem to be particularly prone to uncorrectable errors due to their continuum nature, error-correction techniques have been developed for QCVs \cite{braunstein2003error,ralph2011quantum,van2010note,lloyd1998analog,gottesman2001encoding}, and it is known that fault-tolerant computation is possible via a logical encoding of qudits or qubits inside a universal QCV quantum computer \cite{gottesman2001encoding,menicucci2014fault}. Hence, in addition to qubits, higher dimensional qudits and QCVs also potentially provide a viable route towards a universal, scalable quantum computer. We will now use the term \emph{quantum variable} (QV), as introduced in \cite{Proctorthesis2016,Proctorthesis2016} to refer simultaneously to qubits, higher dimensional qudits or QCVs.

Decoherence is the major obstacle to realising a useful quantum computer, and in order to minimise its destructive effects it is essential that each QV in the `register' of a quantum computer is isolated as effectively as possible. One method for doing this is to require no direct interactions between register QVs and to mediate the necessary entangling gates on the register via some ancillary systems, which are potentially of a different physical type, that are optimised for this purpose. Such `ancilla-based' quantum computation schemes have been extensively developed, e.g., see \cite{spiller2006quantum,munro2005efficient,proctor2014quantum,anders2010ancilla,kashefi2009twisted,proctor2014minimal,proctor2013universal,shah2013ancilla,halil2014minimum,proctor2016hybrid}. 
However, the literature to date largely considers a qubit-based quantum computer (see \cite{roncaglia2011sequential} for an important exception), and as we have outlined above it is not yet clear whether qubits will, or should, be the preferred basic building block of any future quantum computer. In this paper we present ancilla-based gate methods for quantum computation with general QVs, meaning that the models herein can be applied to qubits, higher dimensional qudits and QCVs.

Ancilla-based computational models explicitly allow for a physical implementation in which each element in the register is a `quantum memory' that is well-isolated and tailored towards long coherence times, with more easily manipulated ancillary systems providing the control. Continuing in this line of thought, it is well-motivated to consider ways in which the access needed to the register, in order to implement universal quantum computation on it, can be further reduced to a minimum. The minimum access needed is the ability to perform a single fixed ancilla-register interaction gate (between ancilla-register pairs). It is furthermore natural to minimise the number of interactions required between an ancilla and register elements to implement an entangling gate. The minimum is obviously a single interaction of the ancilla with each register QV, and this is only possible with the aid of measurements of the ancillas to disentangle the ancilla and the register. These controls, along with ancillas prepared in a fixed state, will be all that is required for universal quantum computation in the main model we present herein, which we introduce in Section~\ref{Dadqc}.  To be clear, the model will be able to implement universal quantum computation using only
\begin{enumerate}
\item  A fixed ancilla-register interaction gate.
\item Local measurements of ancillas.
\item Ancillas prepared in a fixed state.
\end{enumerate}

This model can be understood as an extension, to the setting of general QVs, of the qubit-based \emph{ancilla-driven quantum computation} (ADQC) model proposed in \cite{anders2010ancilla,kashefi2009twisted}.  For this reason, the same name will be used for the more general model developed here. Furthermore, we note that the basic gate methods used in our general QV ADQC model are closely related to those proposed by Roncaglia \emph{et al.} \cite{roncaglia2011sequential}, although Ref.~\cite{roncaglia2011sequential} only explictly considers QCVs and qubits.

Arbitrary local measurements of the ancillas may well be challenging in practice, particularly in the case of QCVs. Hence, we will discuss sets of measurements which are sufficient to guarantee that the model may implement universal quantum computation. In particular, it will be shown that for the QCV-based model, with ancillas realised as optical states, homodyne detection and photon-number counting on these ancillas is sufficient for universality. However, in some physical settings implementing more than one type of measurement (or indeed any measurements) on the ancillas may be challenging. For example, if high-fidelity unitary control of the ancillas is not possible and only one observable may be measured. Hence, in Section~\ref{Smincont} we propose a globally unitary model based on a swap-like gate that requires only this fixed interaction, along with ancillas prepared in states from a fixed orthonormal basis, for universality. We begin in Section~\ref{Squdits} with a review of the necessary details of the `quantum variable' formalism \cite{Proctorthesis2016,Proctorthesis2016}.

\section{general quantum variables \label{Squdits}}

In this section the general \emph{quantum variable} (QV) formalism used throughout this paper is introduced and the relevant quantum computation material needed herein is reviewed. The QV formalism is a method for considering qudits of arbitrary dimension (including qubits) and quantum continuous variables (QCVs) simultaneously. This is presented in greater detail by one of these authors in \cite{Proctorthesis2016,Proctorthesis2016} and only the essential material for what follows is covered here. 

A \emph{qudit} is a quantum system with a Hilbert space of finite dimension $d\in \mathbb{Z}$ (with $d\geq 2$). The integers modulo $d$ plays a crucial role for qudits, i.e., the set $\{0,1, \dots, d-1\}$ along with modular arithmetic \cite{vourdas2004quantum} -- we denote this ring $\mathbb{Z}(d)$. A QCV is a quantum system with a continuous degree of freedom taking values in $\mathbb{R}$, i.e., one-dimensional wave mechanics, described by the $\hat{x}$ and $\hat{p}$ operators obeying $[\hat{x},\hat{p}]=i$. We \emph{define} the dimensionality constant, $d$, for QCVs to be $d=2\pi$. It is useful to define the ring $\mathbb{S}_d$, for a general QV, by
\begin{equation}
\mathbb{S}_d :=  \begin{cases}
    \mathbb{Z}(d)       & \quad \text{for a } d\text{-dimensional qudit},  \\
\mathbb{R}  & \quad\text{for a QCV}. \\
  \end{cases}
\end{equation}

For qudits the group of all the $n$-qudit unitaries, $U(d^n)$, is important in quantum computation, but for QCVs it is conventional to only consider the subset of $n$-QCV unitaries containing all operators of the form $U=\exp( i \text{poly}(\hat{x}_k,\hat{p}_k))$ where $\text{poly}(\hat{x}_k,\hat{p}_k)$ is an arbitrary finite-degree polynomial (over $\mathbb{R}$) of the position and momentum operators of all $n$ QCVs \cite{Proctorthesis2016,Lloyd1999quantum}. For notational simplicity, denote this set by $U((2\pi)^n)$, so that in all cases the relevant set of unitaries for quantum computation is $U(d^n)$.

\subsection{The Pauli operators}
For all types of QVs, a \emph{computational basis} may be chosen for the relevant Hilbert space, $\mathcal{B}:=\{\ket{q} \mid q \in \mathbb{S}_d \}$, with basis states obeying $\expect{q|q'} = \delta(q-q')$ where $ \delta(q-q')$ is the Dirac delta function for QCVs and the kronecker delta for qudits (e.g., for QCVs this basis may be defined as the generalised eigenstates of $\hat{x}$). Using this basis we may define the \emph{Fourier gate}, $F$, by 
\begin{equation}
 F\ket{q}:= \frac{1}{\sqrt{d}} \sum_{q' \in \mathbb{S}_d} \omega^{qq'} \ket{q'},
 \end{equation}
with $\omega : = \exp(2 \pi i /d)$ \cite{Proctorthesis2016}. The $\sum_{q' \in \mathbb{S}_d}$ notation denotes that the summation of $q'$ is over all values in $\mathbb{S}_d$, e.g., it is an integral over $\mathbb{R}$ for QCVs. It is easily shown that $F^4 = \mathbb{I}$. For a qubit the Fourier gate is the well-known Hadamard operator (normally denoted $H$).  

A \emph{conjugate basis}, $\mathcal{B}_{+}$, may be defined to contain the orthornormal states $\ket{+_q} := F \ket{q} $ for $ q \in \mathbb{S}_d$, with this notation borrowed from that in common usage for qubits. It is simple to confirm that
   \begin{equation}
    \expect{q | +_{q'}} = \frac{ \omega^{qq'}}{\sqrt{d}}  \hspace{0.5cm} \forall \, q,q' \in \mathbb{S}_d. \label{overlapcomcon}
    \end{equation}

The (generalised) \emph{Pauli operators} are the $q' \in \mathbb{S}_d$ parameterised unitaries defined by 
\begin{equation} 
Z(q') \ket{q}  := \omega^{qq'} \ket{q} , \hspace{1cm} X(q') \ket{q}  := \ket{q+q'} ,  \label{Paulicomp} 
\end{equation}
for all $q,q' \in \mathbb{S}_d$, where the arithmetic is as appropriate for $\mathbb{S}_d$, as should be assumed for all arithmetic in the following unless otherwise stated \cite{Proctorthesis2016}. For qubits these unitaries reduce to (powers of) two of the ordinary Pauli operators. It will be notationally convenient to let $X\equiv X(1)$ and similarly for all other parameterised unitaries (e.g., $Z \equiv Z(1)$).

It may be easily confirmed that the action of the Pauli operators on the conjugate basis is
\begin{equation}
 X(q') \ket{+_{q}} = \omega^{-qq'} \ket{+_q} ,\hspace{0.5cm}  Z(q') \ket{+_q} = \ket{+_{q+q'}} \label{Pauliconj},
 \end{equation}
 for all $q,q' \in \mathbb{S}_d$. Hence, the computational and conjugate bases are eigenstates of $Z(\cdot)$ and $X(\cdot)$ respectively. It will be useful to define the general QV Hermitian `position' and `momentum' operators
 \begin{equation} 
 \hat{x} := \sum_{q\in\mathbb{S}_d} q\ket{q}\bra{q},\hspace{1cm} \hat{p} := \sum_{q\in\mathbb{S}_d} q\ket{+_q}\bra{+_q},
 \end{equation}
 which for QCVs are the standard position and momentum operators.

\subsection{The Pauli and Clifford groups}
The $X(q)$ and $Z(q')$ operators commute up to a phase, specifically:
\begin{equation} 
Z(q) X(q') = \omega^{qq'} X(q') Z(q), \label{Weyl} 
\end{equation}
 for all $q,q' \in \mathbb{S}_d$. Hence, the Pauli operators may be used to define a subgroup of $U(d^n)$. The ($n$-QV) Pauli group, denoted $\mathcal{P}$, is defined to consist of all operators of the form
\begin{equation}
 p_{\xi, \vec{q}} : = \omega^{\xi/2} X(q_1) Z(q_{n+1})  \otimes  .... \otimes X(q_{n}) Z(q_{2n}) ,
 \end{equation}
where $\vec{q}=(q_1,\dots,q_{2n}) \in \mathbb{S}_d^{2n}$ and $\xi \in\mathbb{S}_D$ where  $\mathbb{S}_D=\mathbb{Z}(2d)$ for qudits and $\mathbb{S}_D=\mathbb{R}$ for QCVs \cite{Proctorthesis2016}. This reduces to the well-known qubit Pauli group for $d=2$ (which contains operators with $\mathbb{I}$, $X$, $Z$ or $Y=iXZ$ gates on each qubit along with a global phase factor of $+1,-1,+i$ or $-i$ \cite{gottesman1999heisenberg}) and the Heisenberg-Weyl group for QCVs \cite{bartlett2002efficient}. 

Note that taking $\mathbb{S}_D=\mathbb{Z}(D)$ with $D=2d$ for qudits, rather than setting $D=d$, is only necessary to obtain the desired properties of the Pauli group for even $d$ \cite{farinholt2014ideal}. However, it is perhaps most convenient to take the convention whereby $D$ is always $2d$ \cite{hostens2005stabilizer}, as we do here. Similarly, when Pauli operators are composed we have $p_{\xi,\vec{q}} \, p_{\zeta,\vec{p}}=p_{\xi+\zeta+2\delta,\vec{q}+\vec{p}}$ where $\delta = q_{1}p_{n+1} + q_{2}p_{n+2} + \dots$, and so for qudits it is perhaps ambiguous as to whether to calculate $\delta$ using  modulo $d$ or $2d$ arithmetic. However, as $\omega^{\delta}$ is invariant under changing this convention, the choice is essentially irrelevant.

  The ($n$-QV) Clifford group is defined in terms of the Pauli group by \cite{Proctorthesis2016}
\begin{equation}
 \mathcal{C}: = \{ U \in U(d^n) \mid U  p U^{\dagger} \in \mathcal{P} \hspace{0.2cm} \forall   p  \in \mathcal{P} \}, \end{equation}
which are the unitaries which transform Pauli operators to Pauli operators under conjugation. The Fourier and Pauli operators are Clifford gates and a further important single-QV Clifford gate is the \emph{phase gate}, denoted $P(p)$, defined by 
\begin{equation} 
P(p) \ket{q}:= \omega^{\frac{pq}{2}(q+\varrho_d )} \ket{q}, \label{phasegate}
 \end{equation}
with $p\in\mathbb{S}_D$ and $\varrho_d=1$ for odd-dimension qudits and $\varrho_d=0$ otherwise. The $d$-dependent $\varrho_d$ parameter is required to guarantee that the phase gate has equivalent properties in all dimensions. For qubits the phase gate reduces to $P= \ket{0}\bra{0}+i \ket{1} \bra{1}$. 

An important two-QV Clifford gate is the controlled-$Z$ gate, denoted $\textsc{cz}$ and defined by
\begin{equation} 
\ket{q}\ket{q'} \xrightarrow{ \mathmakebox[0.8cm]{\textsc{cz}}} \omega^{qq'}\ket{q}\ket{q'}.
\end{equation}
This gate acts symmetrically on the QVs.

The $\textsc{cz}$, $F$, $P(p)$ gates and the Pauli operators form a set of generators for the Clifford group, specifically:
\begin{equation}
\mathcal{C} = \langle \textsc{cz},F, P(q), Z(q) \rangle \,\,\,\,\text{with}\,\,\,\, q\in \mathbb{S}_d, \label{Cliffgen} 
\end{equation} 
meaning that any Clifford gate can be decomposed into multiplicative and tensor products of these four gates \cite{Proctorthesis2016,hostens2005stabilizer,farinholt2014ideal,bartlett2002efficient}. For qudits we may set $p=1$ and $q=1$, as obviously $P(p)$ and $Z(q)$ can be obtained by $p$ and $q$ applications of $P$ and $Z$ respectively. It may be directly confirmed that \cite{Proctorthesis2016}
\begin{align}   
  p_{\xi, (q_1,q_2)}   &\xrightarrow{ \mathmakebox[1.2cm]{F}} p_{\xi-2q_1q_2,(-q_2,q_1)}, \label{Eq:conj-F} \\
 p_{\xi ,(q_1,q_2)}  &\xrightarrow{  \mathmakebox[1.2cm]{P(p)} } p_{\xi+pq_1(q_1+\varrho_d) ,(q_1,q_2+pq_1)} , \label{Eq:conj-P} \\ 
p_{\xi ,(q_1 ,q_2 ,q_3,q_4)} &\xrightarrow{ \mathmakebox[1.2cm]{\textsc{cz}} }    p_{\xi +2q_1q_2,(q_1,q_2,q_3+q_2,q_4+q_1)} \label{Eq:conj-CZ}, \end{align}
where $U \xrightarrow{  \mathmakebox[0.4cm]u} U'$ for operators $u$ and $U$ denotes that $uUu^{-1}=U'$.

\subsection{Universal quantum computation \label{Uniqudit}}
An $n$-QV universal quantum computer (UQC) is defined to be a device which can approximate to arbitrary accuracy any unitary operator in $U(d^{n})$ on $n$ QVs \cite{Proctorthesis2016,Lloyd1999quantum,brylinski2002universal}. A quantum computer which can implement \emph{any} two-QV entangling gate along with a set of single-QV gates that can approximate (to arbitrary accuracy) any single-QV gate is universal \cite{Proctorthesis2016,Lloyd1999quantum,brylinski2002universal}. Although the \textsc{cz} gate will be the most important two-QV entangling gate herein, we will also at times require more general controlled-$u$ gates, denoted $C^c_tu$ and defined by
\begin{equation} 
\ket{q}_c\otimes \ket{q'}_t \xrightarrow{ \mathmakebox[0.8cm]{C^c_tu}} \ket{q}_c\otimes u^q\ket{q'}_t,
\end{equation}
for some unitary $u$. Note that this definition is valid even when the two systems are QVs of different types. The super and subscripts will be dropped from the notation when no confusion will arise.

In order to obtain simple universal gate set constructions, an important class of single-QV operators are the rotation gates. The $R(\vartheta)$ rotation gate takes a function parameter, $\vartheta: \mathbb{S}_d \to \mathbb{R}$, and is defined by
\begin{equation}
 R(\vartheta)  \ket{q} := e^{ i \vartheta(q)} \ket{q} .\label{Zdt}
\end{equation}
For all types of QV, some set of rotation gates along with the Fourier gate are a universal set for single-QV gates \cite{Proctorthesis2016,Lloyd1999quantum,zhou2003quantum} and hence such a set along with an entangling gate is sufficient for UQC.

From a practical perspective, it will also be useful to have more specific universal gate sets. It is well-known that computations using only gates from the Clifford group are not universal and are efficiently classically simulatable when QVs are only measured and prepared in the computational basis \cite{bartlett2002efficient,hostens2005stabilizer,de2013linearized,van2013efficient,gottesman1999fault}. However, for prime dimension qudits the addition of any non-Clifford gate to a set of generators for the Clifford group is sufficient for universality \cite{campbell2012magic,nebe2006self,nebe2001invariants} and for QCVs the addition of any continuous power of a non-Clifford gate is sufficient for universality \cite{Proctorthesis2016,Lloyd1999quantum}. In these cases the gate normally considered is a so-called \emph{cubic phase gate} of some sort, which may be defined in general by
\begin{equation} 
D_{3}(q') \ket{q} := \omega^{q^3q'/c} \ket{q}, 
\label{cubic-phase}
\end{equation}
for $q' \in \mathbb{S}_d$ and some suitable constant $c$. 

For all (prime) dimensions of qudit we may take $c=d^3$ and this is the basic generalisation of the well-known `$\pi/8$-gate' for qubits \cite{Proctorthesis2016}, for prime $d>3$ qudits $c=1$ also provides a non-Clifford gate \cite{campbell2014enhanced}, and for QCVs $c=3$ is conventional \cite{gu2009quantum} (in this case the value of $c$ is essentially irrelevant). For non-prime dimension qudits the addition of any $R(\vartheta)$ gate for a `generic' fixed $\vartheta$ to the Clifford group generators is sufficient for universality \cite{Proctorthesis2016}.

\section{Ancilla-driven quantum computation for general quantum variables \label{Dadqc}}
We now present a model of deterministic universal ancilla-based quantum computation which requires a minimal number of ancilla-register interactions per gate and uses only
\begin{enumerate}
\item A fixed ancilla-register interaction gate.
\item Local destructive measurements on individual ancillas.
\item Ancillas prepared in the state $\ket{+_0}$.
\end{enumerate}
This model will be applicable to \emph{all} types of quantum variables, but for now it is convenient to consider only the case in which the register and ancillary QVs are of the same type (i.e., they are all QCVs or all qudits of the same dimension) -- this restriction will be relaxed in Section~\ref{Ascz}. 

It is clearly necessary to carefully choose the ancilla-register interaction, as universal quantum computation will not be possible in this fashion with just any fixed two-QV gate (e.g., obviously it must be entangling). We will initially consider the interaction gate
\begin{equation}
E_{ar} := F_r  F^{\dagger}_a \textsc{cz},
\end{equation}
with alternative interactions discussed in Section~\ref{Ascz}. That is, the model allows the application of $E_{ar}$ to any ancilla-register pair. Note that here (and throughout) the subscript $a$ is used to refer to an ancillary QV and other subscripts will be used to refer to register QVs. 

From a practical perspective it is important to use only a physically plausible set of measurements on the ancillas, and not all local measurements are equally difficult in practice. However, the allowed measurements will not be restricted at this point -- measurements sets that are sufficient for universality will be discussed in Section~\ref{ADQC-measurements}. As already noted, the general QV model proposed here includes, as the qubit special case, the \emph{ancilla-driven quantum computation} (ADQC) model proposed by Anders \emph{et al.} \cite{anders2010ancilla,kashefi2009twisted} (up to a very minor alteration, noted later). For this reason, the same name is used here.

\subsection{A universal gate set}
Universal quantum computation can be implemented in this general quantum variable ADQC model as follows. It is simple to confirm that the action of the fixed interaction, $E_{ar}$, on a register QV in the state $\ket{q}$ and an initialised ancilla is
\begin{equation} 
  \ket{q}_r \ket{+_0}_a \xrightarrow{E_{ar}}   \ket{+_q}_r \ket{q}_a. 
  \label{delocal} 
  \end{equation}
Hence, an interaction of a register QV with an ancilla delocalises a logical QV in the register over the two physical QVs. Therefore, any subsequent manipulations (i.e., gates or measurements) on the ancilla will implement transformations on the logical QV, and a measurement of the ancilla will destroy this delocalisation. It is this delocalisation which enables the following universal gate set implementation.

Sequential interactions between an ancilla and two register QVs, $r$ and $s$, followed by an $\hat{x}$ measurement  of the ancilla (often termed a \emph{computational basis measurement}) implements an entangling gate on this pair of register QVs  \footnote{Here there is one minor difference between the model herein in the qubit sub-case and the qubit-based model of \cite{anders2010ancilla} and \cite{kashefi2009twisted}. Specifically, Anders \emph{et al.} use a measurement of the Pauli operator $Y:=iXZ=i(\ket{1}\bra{0}-\ket{0}\bra{1})$. Implementing the entangling gate in the general QV model herein with a measurement constructed from the eigenstates of $Y:=\omega^{(1+\varrho_d)/2}XZ$ (defined for a general QV) is possible. However, the mathematical details are substantially more complicated and as this is unnecessary it is therefore avoided.}. This is because
\begin{equation}
 \ket{q}_r|q'\rangle_s \ket{+_0} \xrightarrow{E_{as} E_{ar}}  \omega^{qq'} \ket{+_q}_r\ket{+_{q'}}_s\ket{+_{-q}}, \end{equation} 
and the $\omega^{qq'}$ phase is exactly the phase that would be created by a $\textsc{cz}$ gate acting on these two register QVs. Therefore, given that the $\hat{x}$ measurement outcome is $m\in \mathbb{S}_d$, the gate implemented after the ancilla has been measured may be confirmed to be
\begin{align} 
\frac{ \bra{m}_a E_{as} E_{ar}  \ket{+_0}_a}{\|\bra{m}_a E_{as} E_{ar}   \ket{+_0}_a\| }
&=  X_r(m) \tilde{E}_{rs}, \label{ADQCent}
 \end{align} 
where $\tilde{E}_{rs}$ is the symmetric entangling gate given by 
\begin{equation} \tilde{E}_{rs} =F_r F_s  \textsc{cz}. \label{ADQCent2}
\end{equation} 
Hence, an entangling gate has been implemented up to a measurement outcome-dependent Pauli error gate, $X(m)$.

This may be summarised in the quantum-classical circuit diagram
  
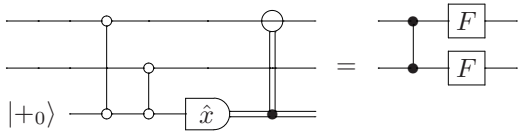
\begin{equation*}
  \Qcircuit @C=1.2em @R=0.5em {
&  \qw&  \qw & \ctrlo{2} &   \qw   & \qw &\targminus & \qw  & & & \ctrl{1} & \gate{F} &  \qw  \\
&  \qw &\qw & \qw  &  \ctrlo{1}   & \qw & \cwx \qw & \qw  &  = & & \control \qw &  \gate{F} & \qw  \\
&& \lstick{\ket{+_0}} & \controlo \qw & \controlo \qw   &  \measureD{\hat{x}} &\control \cwx \cw   & \cw    & & &  &  
 }  
 \end{equation*}
which includes an explicit correction for the $X(m)$ error -- such a correction is \emph{not} required for deterministic universal quantum computation, as will be seen later. Note that in this diagram two quantum wires connected via a line and `$\circ$' symbols denotes the fixed ancilla-register interaction, quantum wires joined by a line and `$\bullet$' symbols is the standard notation for $\textsc{cz}$, the double lines represent a classical variable of the apprioriate type (e.g., a bit, dit or CV) and the `$\ominus$' is a natural notation for a $X^{\dagger}$ gate -- as this is a subtraction gate.

A $FR(\vartheta)$ gate, for any $\vartheta$ phase function, can be implemented on a register QV (up to a Pauli error) by interacting the register QV with an ancilla and then performing a $\vartheta$-dependent measurement on the ancilla. The specific measurement is of $\hat{x}_{FR(\vartheta)}$, where this uses the shorthand
\begin{equation}
\hat{x}_{u} := u^{\dagger} \hat{x} u = \sum_{q \in \mathbb{S}_d} q  \left(u^{\dagger} \ket{q}\bra{q} u \right)   .
\end{equation}
This may be confirmed by showing that
\begin{equation}
  \frac{ \bra{m}F_aR_a(\vartheta) E'_{ar}  \ket{+_0}}{\|\bra{m}F_aR_a(\vartheta) E'_{ar}   \ket{+_0}\| } = X_r(-m) F_rR_r(\vartheta),
   \label{ADQCvtheta}
  \end{equation}
where $m \in\mathbb{S}_d$ is the measurement outcome. This gate method is summarised in the quantum-classical circuit diagram
 
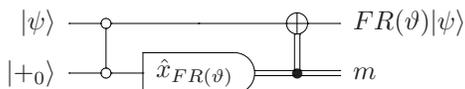
\begin{equation*}
  \Qcircuit @C=1.2em @R=0.7em {
 \lstick{\ket{\psi}} & \ctrlo{1}  & \qw  &\targ  &   \rstick{FR(\vartheta)\ket{\psi}}   \qw \\
\lstick{\ket{+_0}} & \controlo \qw    &  \measureD{\hat{x}_{FR(\vartheta)} } &\control \cwx \cw   &    \rstick{m}  \cw 
 }  
 \end{equation*}
which again explicitly corrects for the error. Note that this gate method is essentially equivalent to that proposed in Ref.~\cite{roncaglia2011sequential}. 

Ignoring the Pauli errors for now, the two gate methods presented above implement gates which are sufficient for universal quantum computation for all types of QVs as they can generate an entangling gate, $F$ (by taking the phase function to be $\vartheta(q)=0$ for all $q$) and any rotation gate (as $F^3FR(\vartheta)=R(\vartheta)$).

\subsection{Step-wise determinism}
The Pauli errors may be accounted for using classical feed-forward of measurement outcomes and adaptive measurements, instead of using explicit local correction gates which are not available in the ADQC model. This is directly analogous to the techniques of measurement-based quantum computation \cite{raussendorf2001one,raussendorf2003measurement,zhou2003quantum,menicucci2006universal} which have been recently presented in the general quantum variable formalism used here in Ref.~\cite{proctor15measurement}. 

Consider an $n$-QV computational register and write the state it is in as $ p_{\zeta,\vec{q}}  \ket{\psi}$, where $p_{\zeta,\vec{q}}$ is any Pauli operator. It is convenient to write $\vec{q}=(x_1,\dots,x_n,z_1,\dots,z_n)$, as then the error on the $k$\textsuperscript{th} QV is $X_k(x_k)Z_k(z_k)$. Given the vector $\vec{q}$, we may implement the mapping $p_{\xi,\vec{q}}  \ket{\psi} \to p_{\eta,\vec{p}} U \ket{\psi}$, where $U=FR(\vartheta)$ or $U=\tilde{E}$ on any QV(s), using the available operations in ADQC (and so without explicit local corrections). Repeated applications of these processes allow for a deterministic implementation of any quantum computation up to final Pauli errors on each QV, which can then be accounted for in classical post-processing of final measurement outcomes. Note that the natural way to think of $\vec{q}$ is as $2n$ classical variables on which classical computations are implemented in parallel to the quantum computation on the $n$ QVs.

 The entangling gate $\tilde{E}$ is a Clifford gate, and hence $X_r(m)\tilde{E}_{rs} p_{\xi,\vec{q}} = p_{\eta,\vec{p}} \tilde{E}_{rs}$  for some $\eta$ and $\vec{p}$. Hence, to implement a $\tilde{E}$ gate on a register with (possible) Pauli errors, no adaptive element needs to be added to the process in Eq.~(\ref{ADQCent}) and it is only necessary to implement a classical computation to update $\vec{q} \to \vec{p}$. The global phase is irrelevant, so we need not compute $\xi \to \eta$. It is simple to confirm (using Eq.~(\ref{Eq:conj-F}) and Eq.~(\ref{Eq:conj-CZ})) that the classical computation required is
\begin{equation} 
(x_r,x_s,z_{r},z_{s}) \xrightarrow{} (m-z_{r}-x_s,-z_{s}-x_r,x_r,x_{s}).
\end{equation}
This can be achieved with classical $\textsc{sum}$, $\textsc{swap}$ and inversion ($x\to-x$) gates (as always, $-x$ is taken modulo $d$ for dits).

To clarify this process, it may be written as a quantum-classical circuit which acts on two register QVs, one ancillary QV and four classical variables. Specifically, this process to implement $\tilde{E}$ and update the classical variables is summarised with the circuit  
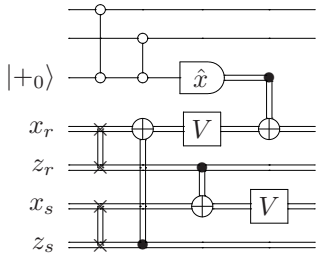
\vspace{-0.3cm}
  \[
  \Qcircuit @C=1.0em @R=0.7em {
& \ctrlo{2} & \qw &   \qw & \qw &\qw  \\
& \qw & \ctrlo{1} &   \qw &  \qw & \qw  \\
 \lstick{\ket{+_0}} & \controlo \qw   & \controlo \qw  &  \measureD{\hat{x}} &\control  \cw  &  \\
 \lstick{x_r} & \cswap           & \ctarg \cw  & \cgate{V}\cw  &\ctarg \cwx \cw    &\cw   \\ 
 \lstick{z_r}&\cswap \cwx  &\cwx \cw   & \control \cw & \cw  & \cw      \\ 
 \lstick{x_s}&\cswap           &  \cwx \cw &\ctarg \cwx \cw   &\cgate{V}  \cw   & \cw    \\ 
 \lstick{z_s}&\cswap \cwx  & \control \cwx \cw  &   \cw   & \cw   & \cw    
 }  \vspace{0.2cm}
 \]
 where the first and second quantum wires represent the $r$ and $s$ QVs respectively, $V$ denotes the inversion operator $x\to-x$, and wires connected via a line and `$\times$' symbols is the standard notation for the $\textsc{swap}$ gate, which maps $(x,z)\to(z,x)$. 

To apply a $FR(\vartheta)$ gate on one of the QVs that has Pauli errors, the measurement used in the process of Eq.~(\ref{ADQCvtheta}) must (in general) be classically adapted. The $X(x)$ gate maps $\ket{q} \to \ket{q+x}$. Hence, defining $\vartheta_{x}$ to be the phase function given by $\vartheta_{x}(q)= \vartheta(q+x)$, it follows from Eq.~(\ref{Eq:conj-F}) that
\begin{equation*} 
 X(-m)FR(\vartheta_{x}) X(x)Z(z) = \omega^{-xz}X(-z-m)Z(x)    FR(\vartheta).
 \label{CV-adapted}
 \end{equation*} 
Therefore, to implement a $FR(\vartheta)$ gate on the $r$\textsuperscript{th} QV, the measurement of the ancilla after it interacts with the register QV should be of the $x_r$-adapted operator $\hat{x}_{FR(\vartheta_{x_r})}$, which implements $X(-m)FR(\vartheta_{x_r})$ on the register with $m$ the measurement outcome. The corresponding update of the classical variables is
\begin{equation}
 (x_r,z_r) \xrightarrow{} (-z_r-m,x_r).
\end{equation}

This may be written as the quantum-classical circuit module  
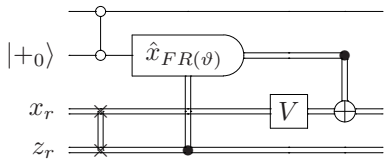
\vspace{-0.2cm}
  \[
  \Qcircuit @C=1.0em @R=0.7em {
 & \ctrlo{1} &   \qw &  \qw & \qw & \qw  \\
 \lstick{\ket{+_0}}  & \controlo \qw   & \measureD{\hat{x}_{FR(\vartheta)}}& \cw &\control  \cw  &  \\
 \lstick{x_r} & \cswap           &  \cwx \cw  & \cgate{V}\cw  &\ctarg \cwx \cw    &\cw   \\ 
 \lstick{z_r}&\cswap \cwx  &  \control \cwx \cw   &\cw & \cw  & \cw      \\
 } \vspace{0.1cm}
 \]
 where the adaption to the measurement basis is shown schematically via the classical control wire to the measurement device.
 
When the $FR(\vartheta)$ operator is a Clifford gate the measurement dependency can be removed from this procedure at the cost of further classical computation. The error update procedure for the $FP(p)$ gate on the $r$\textsuperscript{th} QV when no classical control is used can be found from Eqs.~(\ref{Eq:conj-F} -- \ref{Eq:conj-P}) to be
 \begin{equation}
    (x_r,z_r) \xrightarrow{FP(p)} (-z_r-px_r-m,x_r).
\end{equation}
Written as a quantum-classical circuit module, the $FP(p)$ gate may be implemented by 
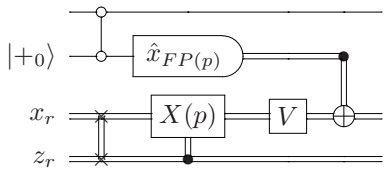
\vspace{-0.2cm}
  \[     \Qcircuit @C=1.0em @R=0.7em {
 & \ctrlo{1} &   \qw &  \qw & \qw & \qw  \\
 \lstick{\ket{+_0}}  & \controlo \qw   & \measureD{\hat{x}_{FP(p)}} & \cw &\control  \cw  &  \\
 \lstick{x_r} & \cswap           &  \cgate{X(p)} \cw  & \cgate{V}\cw  &\ctarg \cwx \cw    &\cw   \\ 
 \lstick{z_r}&\cswap \cwx  &\control \cwx \cw   &\cw & \cw  & \cw      \\
 }  \vspace{0.1cm}
 \]
Finally, note that $Z(q)$ and $X(q)$ gates can be implemented with only classical processing. That is, to implement a $X_r(q)Z_r(q')$ gate simply map the classical variables for the $r$\textsuperscript{th} QV as $(x_r,z_r) \to (x_r-q, z_r-q')$. 

Because $\textsc{cz}$, $FP(p)$ and $Z(q)$ are sufficient to implement any Clifford gate (see Eq.~(\ref{Cliffgen})) and methods for implementing these operators have been given which require no classically-adapted measurements, then no measurement dependencies are required for any Clifford gates. This is not surprising, given the close relation of this model to MBQC for general QVs, which will become particularly clear in Section~\ref{SlinkADtoMB}.

\subsection{Universal sets of measurements \label{ADQC-measurements}}

The gate methods given so far are sufficient for universal quantum computation on the register. However, these techniques include $\hat{x}_{FR(\vartheta)}$ measurements for unspecified phase-functions $\vartheta:\mathbb{S}_d \to \mathbb{R}$, and not all such measurements will be equally straightforward in practice. In order to implement any Clifford gate a very limited set of measurements is required. For qudits only three measurement operators are necessary: $\hat{x}$, $ \hat{x}_F $ and $\hat{x}_{FP}$ (as $F$, $FP$, $Z$ and $\tilde{E}$ generate the Clifford group for qudits \cite{Proctorthesis2016,farinholt2014ideal}).  For a qubit, these are equivalent to measurements of the Pauli $Z$, $X$ and $Y$ operators, respectively, up to a post-processing on the measurement outcomes of $+1\to 0$, and $-1\to 1$. 

For QCVs, any Clifford gate can be implemented via measurement of the quadrature operator $X(\phi) = \hat{p}  \cos \phi + \hat{x} \sin \phi$ for variable $\phi\in [0,2\pi )$, although this must be augmented with additional post-processing on the measurement outcomes. This is because $\hat{x}=X(\pi/2)$ and $\hat{x}_{F P(\tan\phi)}= - X(\phi)/\cos\phi$, which may be shown using the $k=2$ case of the QCV relations
 \begin{equation}
\hat{x}  \xrightarrow{ D_k(q) } \hat{x}, \hspace{1cm}  \hat{p}  \xrightarrow{ D_k(q) } \hat{p} - q \hat{x}^{k-1},
\label{Dk-conj}
\end{equation}
  where $D_{k}(q)\ket{q} = \omega^{ q \hat{x} ^k /k}\ket{q}$. Quadrature measurements (also termed homodyne detection) are now routine in quantum optics, e.g., see Ref.~\cite{ukai2011demonstration,su2013gate}. Although the most natural realisation of the ancillary systems in QCV-based ADQC is probably an encoding into optical states, interestingly, homodyne detection of QCVs encoded into atoms has also recently been demonstrated \cite{gross2011atomic}.

As discussed in Section~\ref{Uniqudit}, a non-Clifford gate is necessary for universal quantum computation. For qudits, there is no obvious \emph{physical} reason why one gate in particular should be picked to obtain universality and there are a large range of fixed measurements which would suffice in conjunction with the Clifford measurements. In a given physical set-up the easiest such measurement (and its classically-adapted versions) could be chosen. A range of suitable variable-basis measurements, or equivalently, variable local gates followed by a fixed-basis measurement, have been implemented in atomic higher dimensional qudits \cite{neeley2009emulation,anderson2014accurate}, and are common practice in qubit systems, e.g., see Refs.~\cite{lanyon2013measurement,barz2014demonstrating,gao2011experimental}. 

For QCVs, a single-QCV gate is a non-Clifford unitary if and only if it is generated by a Hamiltonian which is at least a cubic function of $\hat{x}$ and $\hat{p}$ \cite{Lloyd1999quantum,bartlett2002efficient}. The natural gate to consider is the cubic phase gate, as introduced in Eq.~(\ref{cubic-phase}). This cubic phase gate (followed by $F$) may be implemented via a measurement of the operator 
\begin{equation}\hat{x}_{FD_3(q) }= q \hat{x}^2 -\hat{p},
\label{D3oper}
\end{equation}
 where this equality follows directly from Eq.~(\ref{Dk-conj}). The adaptive version of this gate required for direct step-wise determinism is simply given by letting $q \to q +x$ in Eq.~(\ref{D3oper}), where $x$ is the classical variable tracking the $X$-type error on the relevant register QV, which is the operator $(q+x)\hat{x}^2-\hat{p}$. This can also be decomposed into a measurement of the operator in Eq.~(\ref{D3oper}) followed by $x$-dependent Clifford gates. The next subsection implicitly covers how to do this.

\subsection{Finite squeezing distortions and cubic phase states for QCVs \label{finite-s}}

There are two difficulties with physically realising the ADQC model which are specific to the setting of QCVs and these are now addressed. Firstly, ideal computational and conjugate basis states (and so the initial ancilla states $\ket{+_0}$) are unphysical \cite{Lloyd1999quantum} and they may only be approximated. Define the (Clifford) squeezing operator by $S(s)\ket{q}:=\ket{sq}$ with $s>0$, which may be also be written as $S(s)=\exp(-i\ln (s) (\hat{x}\hat{p}+\hat{p}\hat{x})/2)$, and let $\ket{\text{vac}}$ denote the vacuum. Then $  S(s) \ket{\text{vac}}\approx \ket{+_0} $ when $s \gg 1$ and $S(s) \ket{\text{vac}}\approx \ket{0} $ when $s \ll1$ \cite{radmore1997methods}. The effect on the QCV ADQC computation of preparing the ancillas in such approximations to $\ket{+_0}$ is the introduction of Gaussian noise to the register with the application of each gate, as can be inferred from Ref.~\cite{gu2009quantum}, in which the effect of such approximations on QCV gate teleportation is analysed. Furthermore, this distortion will build up linearly with the number of gates implemented \cite{gu2009quantum}. Recently it has been shown that in QCV MBQC these errors can be mitigated for by encoding qubits into the QCVs (using the technique of Ref.~\cite{gottesman2001encoding}) as long as the squeezing is above a threshold value \cite{menicucci2014fault}. This threshold is around 20 dB \cite{menicucci2014fault} which is higher than the current experimental record of 12.7 dB \cite{eberle2010quantum,mehmet2011squeezed} (the state $S(s)\ket{\text{vac}}$ has $10 \log_{10} (s^2)$ dB of squeezing \cite{lvovsky2014squeezed}). An extension of this finite-squeezing fault tolerance technique to the QCV ADQC model is left for future work, although it is noted that it is likely that any fault-tolerance threshold would be above the currently experimentally obtainable values.

The second issue with the QCV ADQC model, particularly with optical ancillas, is that the measurement to directly implement the cubic phase gate (and obtain universality) is very difficult to achieve experimentally -- such a measurement requires a non-linear optical element. One alternative to these measurements is to use auxiliary resource states, such as the so-called \emph{cubic phase states} \cite{gottesman2001encoding} and convert these to cubic phase gates.  We now show how this technique can be implemented within the ADQC model.

We first show how to create the state $D_3(\gamma) \ket{+_0}$ using only homodyne detection, photon-number counting (assuming the setting of optical ancillas, as we do for now), and the fixed ancilla-register interaction. In Ref.~\cite{gottesman2001encoding} it is shown how to approximately generate a cubic phase state with Gaussian operations acting on squeezed vacua and a measurement of the number operator, $\hat{n}=(\hat{x}^2+\hat{p}^2-1)/2$. From Refs.~\cite{gottesman2001encoding,gu2009quantum} (in particular, see Eq.~(45) of Ref.~\cite{gu2009quantum}) it may be confirmed that
  \[     \Qcircuit @C=1.0em @R=0.7em {
 \lstick{S(s)\ket{\text{vac}}}& \ctrlo{1} &   \gate{F^{\dagger}} & \qw & \rstick{\approx D_3( \gamma(n)) \ket{+_0}} \qw  \\
 \lstick{S(s)\ket{\text{vac}}}  & \controlo \qw  & \gate{Z(-q)}  & \measureD{\hat{n} } &\rstick{n} \cw   \\
 }  \vspace{0.3cm}
 \]
 where $\gamma(n) =(2\sqrt{2n+1})^{-1}$ and with this approximation holding when $s\gg1$ and $q \gg s$. In this circuit and our setting of QCV ADQC, the lower quantum wire represents an ancilla initialised in an approximation to $\ket{+_0}$ and the top wire represents an auxiliary \emph{register} QCV initialised similarly. Note that the local $F^{\dagger}$ gate on the register QCV may be applied (up to a Pauli error) via an ancilla-driven gate using homodyne detection. The measurement on the ancilla here is a displacement, which is simple experimentally, followed by a photon-number resolving detector (PND). There have been many recent improvements in the state-of-the-art in PNDs \cite{humphreys2015tomography,calkins2013high} and, although such measurements are still highly challenging, they are perhaps currently the most well-developed non-Gaussian optical component.

 Auxiliary register QCVs prepared in cubic phase states, using the method above, can be used to implement cubic phase gates in ADQC, using only homodyne detection and the ancilla-register interaction, as we now show (this is an adaption of a QCV MBQC method presented in Ref.~\cite{gu2009quantum}). Eq.~(\ref{Dk-conj}) implies that
  \begin{align}
  Z(q)  &\xrightarrow{ D_3(\gamma) } Z(q),\\
X(q) & \xrightarrow{ D_3(\gamma) }  e^{i q ( \gamma \hat{x}^2-\hat{p} ) } =: C(q,\gamma),
  \label{DX-DZ}
 \end{align}
where $C(q,\gamma)$ is a Clifford gate as it is generated by a Hamiltonian that is quadratic in $\hat{x}$ and $\hat{p}$. Now, by noting that any diagonal single-QV gate commutes with the control of a controlled gate, it is not hard to confirm that   
  \begin{equation}  \Qcircuit @C=0.8em @R=0.9em {
\lstick{X(x)Z(z) \ket{\psi} } & \targminus &   \measureD{\hat{x} } &\rstick{m} \cw  \\
 \lstick{ D_3(\gamma) \ket{+_0} }  & \ctrl{-1} \qw & \qw & \rstick{C(x',\gamma) Z(z) \ket{ \psi'} } \qw   \\
 }   \vspace{0.3cm}
 \label{circ-1}
 \end{equation}
 where $\ket{\psi'}=D_3(\gamma)\ket{\psi}$ and $x'=x-m$. In this context both of these QCVs represent register QCVs. Hence, by using an auxiliary cubic phase state the cubic phase gate has been implemented on a computational register QCV with pre-existing Pauli errors and in the process the computational QCV has been teleported to the auxiliary QCV and a (non-Pauli) Clifford error has been created, in addition to an ordinary Pauli error.
 
Before discussing the Clifford error, it is important to note that this circuit can be implemented with an ancilla-driven sequence: The $CX^{\dagger}$ gate in this circuit is a Clifford gate (it may be decomposed into $\tilde{E}$ and $F$ gates) and hence it may be implemented via ancillas and homodyne detection. Furthermore, the $\hat{x}$ measurement of the first register QCV may be simulated using the ancilla-driven circuit
 \begin{equation}    \Qcircuit @C=1.2em @R=1.0em {
\lstick{ \ket{\psi} }  & \ctrlo{1} &  \qw &\rstick{\ket{+_m}} \qw \\ 
 \lstick{  \ket{+_0}}  & \controlo \qw & \measureD{\hat{x} } &\rstick{m} \cw \\
 } 
 \label{x-measure}
 \end{equation}
which is equivalent to an $\hat{x}$ measurement on the register QV (more precisely, it is equivalent to a non-destructive measurement of $\hat{x}$ followed by $F$, but we may assume the register QV is discarded).

Finally, the Clifford error in Eq.~(\ref{circ-1}) can be converted to a Pauli error via an ancilla-driven $C(m-q,\gamma)$ gate, as such a gate can be implemented up to Pauli errors by decomposing this Clifford gate into a sequence of $FP(p)$ and $F$ Clifford gates. Hence, we have given a practical method for making  an auxiliary cubic phase state in QCV ADQC and we have shown how it may be used to implement the mapping 
\begin{equation}
 X(x)Z(z)\ket{\psi} \to X(x')Z(z') D_3(\gamma)\ket{\psi},
\end{equation}
for any arbitrary logical register state $\ket{\psi}$, with $\gamma$ fixed by the outcome of the photon number detection. Moreover, this can be converted to a cubic phase gate $D_3(q)$ with any $q\in \mathbb{R}$, by noting that  $D_3(q) = S(\gamma/q)D_3(\gamma)S(q/ \gamma )$, where these squeezing gates may themselves be implemented via homodyne detection on ancillas.

\subsection{Parallel computation in ADQC \label{SlinkADtoMB}}

The ADQC model for general quantum variables has a range of physically appealing properties, as we have already discussed. Moreover, it also has interesting computational properties. In particular, it has the same `parallelism' as MBQC. The qubit MBQC model is well-known to be more powerful than quantum circuits for parallel computation \cite{danos2007measurement,Broadbent20092489,browne2011computational} and the higher dimensional qudit and QCV models have similar properties \cite{zhou2003quantum,menicucci2006universal}. Recently it has been shown that the parallelism inherent in MBQC for all types of QVs is essentially equivalent \cite{proctor15measurement,Proctorthesis2016} and can be understood as providing the ability to implement any Clifford gate in a single layer of quantum computation, which is less than the logarithmic (in $n$) number of quantum layers required to implement an arbitrary $n$-QV Clifford gate in a quantum circuit \cite{proctor15measurement,Proctorthesis2016}. We now explain how the ADQC model has access to the parallel power of MBQC by showing how an MBQC computation can be simulated in ADQC with only a constant increase in the number of computational layers (a more formal proof is presented in \cite{Proctorthesis2016}). 
\newline
\indent
In MBQC with any type of QVs the computation can be broken down into two sequential stages \cite{proctor15measurement}: (1) Layers of $\textsc{cz}$ gates on an initial product state (all QVs, except possibly the input are initialised to $\ket{+_0}$), creating an entangled `resource state'. (2) Layers of $\hat{x}_{FR(\vartheta)}$ measurements, where the $\vartheta$ functions may depend on measurement outcomes in earlier layers. In each `$\textsc{cz}$ layer' at most one \textsc{cz} acts on each QV. Hence, each such layer may be easily simulated in ADQC with only a constant (nine) number of layers. Specifically, $\textsc{cz}= F^3 \otimes F^3 \cdot \tilde{E}$, which requires seven ancillas to implement with ADQC gates and takes no more than nine layers (two layers for each $F$ gate, in parallel, three layers for the $\tilde{E}$ gate), and each such \textsc{cz} gate in a layer may be implemented in parallel. Each layer of measurements may be implemented using no more than four layers of ADQC computation: Each $\hat{x}_{FR(\vartheta)}$ measurement may be decomposed into first a $FR(\vartheta)$ gate, where $\vartheta$ may depend on outcomes from previous layers in the MBQC (and hence previous layers in the ADQC simulation), followed by an $\hat{x}$ measurement. A local $FR(\vartheta)$ gate is easily applied via an ancilla, using the method of Eq.~(\ref{ADQCvtheta}) (which use two layers). An $\hat{x}$ measurement on a register QV can be simulated in ADQC using the procedure of Eq.~(\ref{x-measure}) (which uses two layers). Each such measurement simulation in the layer can be implemented in parallel. 
\newline
\indent
In summary, an MBQC computation can be simulated with only a small constant overhead in the number of quantum computational layers. Note that additional Pauli errors are created in the ADQC simulation of the MBQC, as there are more measurements, but these can simply be absorbed into the classical side-processing. Therefore, ADQC has access to at least the same `parallelism' as MBQC, and their parallel power is actually identical -- this may be confirmed by showing that an MBQC computation can also simulate an ADQC computation with constant overhead, which follows from results in Refs.~\cite{proctor15measurement,Proctorthesis2016}. We have already implicitly seen that the ADQC model can also be used to drive a quantum circuit model (i.e., unitary gates only) computation, and hence ADQC can be understood as a hybrid between the MBQC and quantum circuit models. Interestingly, similar conclusions may also be reached via considering `local complementations' of graphs and the MBQC cluster state formalism, as in Ref.~\cite{roncaglia2011sequential}.

\subsection{Alterations and extensions to the ADQC model \label{Ascz}}
One of the first constraints imposed on the ADQC model herein was that the ancillary and register QVs were all of the same type. It is possible to extend the ancilla-driven model to apply to the `hybrid' setting when the ancillary and register systems are no longer the same type of QVs, as we now show. 

Let $d_a$ and $d$ be the dimensionality constants for the ancillary and register QVs, respectively. In the following, it will be assumed that the register does \emph{not} consist of QCVs, as in that case the relations presented below only hold when the ancillas are also QCVs \cite{Proctorthesis2016} and this case has already been covered above. Consider the natural extension of $E_{ar}$ to this hybrid setting, which is the fixed interaction
\begin{equation} E'_{ar} := F_r F_a^{\dagger} C^r_a Z. 
\end{equation}
Note that the two $F$ gates here are different, in the sense that they are the gates for the appropriate dimensions of the register and ancillary QVs, and the control direction in the `hybrid-\textsc{cz}' gate is explicitly denoted as when the dimension do not match then $C^r_a Z\neq C^a_r Z$ (the $Z$ gates are not the same in each case). 

The natural extension of the entangling gate technique in Eq.~(\ref{ADQCent2}), that is, two register qudits interacting with an ancilla on which $\hat{x}$ is measured, implements the gate
\begin{align} 
\frac{ \bra{m}_a E'_{as} E'_{ar}  \ket{+_0}_a}{\|\bra{m}_a E'_{as} E'_{ar}   \ket{+_0}_a\| }
&= u_r(m) \tilde{E}_{rs}',\label{ADQCentD}
 \end{align} 
where $m$ is the measurement outcome, $\tilde{E}'_{rs}$ is the symmetric entangling gate given by $\tilde{E}_{rs}' =F_r F_s  Cu$ (where $u \equiv u(1)$) and the gate $u(q')$ is defined by the action
\begin{equation} u(q')\ket{+_q} := e^{-2 \pi i qq'/d_a} \ket{+_q}. \end{equation}
Note that $u(\cdot)$ is \emph{not} a Pauli gate, in general. 

Furthermore (extending Eq.~(\ref{ADQCvtheta})) we have that by interacting an ancilla and register QV and measuring the ancilla in the basis $\hat{x}_{FR(\vartheta)}$ (here $\vartheta:\mathbb{S}_{d_a} \to \mathbb{R}$) the gate
\begin{equation}
  \frac{ \bra{m}F_aR_a(\vartheta) E'_{ar}  \ket{+_0}}{\|\bra{m}F_aR_a(\vartheta) E'_{ar}   \ket{+_0}\| } = u_r(-m) F_rR_r(\bar{\vartheta}),
   \label{ADQCvthetaD}
  \end{equation}
is implemented, where $m$ is the measurement outcome and $\bar{\vartheta}$ is the phase-function given by $\bar{\vartheta} (q) = \vartheta(0 \oplus q)$ for $q\in \mathbb{S}_d$ with $\oplus$ denoting the arithmetic of $\mathbb{S}_{d_a}$. 

 Ignoring the $m$-dependent error gates for now, consider the gate set these methods can implement. When $d_a \geq d$ or the ancilla is a QCV, any $FR(\vartheta$) operator may be applied to the register (up to the error) by an appropriate choice of measurement basis for the ancilla (as $0 \oplus q =q$). However, when $d_a < d$ then, no matter what measurement is chosen, the gate implemented has a phase function which obeys $\bar{\vartheta} (q)= \bar{\vartheta} (q \text{ mod } d_a)$. For example, if the ancillas are qubits then each $FR(\bar{\vartheta})$ gate that can be implemented on the register has a phase-function $\bar{\vartheta}$ with $\bar{\vartheta}(q)=\vartheta(0)$ if $q$ is even and $\bar{\vartheta}(q)=\vartheta(1)$ if $q$ is odd for some $\vartheta:\{0,1\} \to \mathbb{R}$, which is fixed by the choice of measurement basis. Therefore, when $d_a \geq d$ it is clear that the gate set is universal (an entangling gate along with all $FR(\vartheta)$ gates) but it is not clear that this is the case for any $d_a<d$. Such a gate set may be universal for some values of $d$ and $d_a<d$, but it seems unlikely that such a gate set is universal in all cases and it would need to be considered on a case-by-case basis.
   
Consider now the $u(\pm m)$ error gates. In order to account for these errors using the step-wise determinism techniques employed herein it is necessary for $\tilde{E}'$ to be Clifford and for $u(\pm m)$ to be a Pauli gate for all measurement outcomes. The condition under which this holds is when $d_a=d/k$ for some positive integer $k$, where $d$ is the dimensionality of the register qudits, as in this case then $u(\pm m)=X(\pm km)$. Hence, we must have ancillary qudits with $d_a \leq d$ for step-wise determinism, but unless $d_a=d$ it may not be possible to implement a universal gate set on the register. Alternatively, when the ancillas are qudits of dimension $d_a>d$ or QCVs, the gate set which may be applied to the register is universal but step-wise determinism is not possible (unless explicit local corrections on the register are available). However, in this setting the model can be said to be universal in a stochastic sense -- any quantum computation can be implemented with a stochastic sequence of single-qudit gates of indeterminate length between each entangling gate (which can be deterministically applied, up to a $u(m)$ error). This is a form of what is termed \emph{repeat-until-success} gate implementation \cite{lim2005repeat,bocharov2014efficient,paetznick2013repeat}, and the properties of computing in this fashion have been discussed in detail elsewhere (e.g., see Refs.~\cite{paetznick2013repeat,halil2014minimum,shah2013ancilla}).

We return now to the setting in which the register and ancillary systems are of the same type. The choice to take the fixed ancilla-register interaction gate to be $E_{ar}=F_rF^{\dagger}_a\C^r_a Z$ was made at the beginning of this section, and it is not obvious that this interaction has unique properties that single it out as the only possible option. Indeed, there is an alternative interaction which allows for deterministic universal quantum computation in ADQC. It is based on the $\textsc{swap}$ gate and is given by
\begin{equation} 
\check{E}_{ar}:= F_a \cdot \textsc{swap} \cdot \textsc{cz},
\label{check-S}
 \end{equation}
 where $\textsc{swap}$ is defined by
\begin{equation}\ket{q}\ket{q'} \xrightarrow{ \textsc{swap}} \ket{q'}\ket{q}.\end{equation} 
When considering this fixed interaction there are some minor changes needed to the gate implementation methods, which are outlined in Appendix~\ref{AAscz}. 

Note that with this interaction the close link between the ADQC model and MBQC (for all types of QVs) is particularly clear: The \textsc{swap} gate in the $\check{E}_{ar}$ interaction entangles and interchanges the register QV with an ancillary QV, and to implement a one-QV gate the ancilla is then measured -- if the \textsc{swap} gate is instead absorbed into the initial state this becomes state teleporation along a two-QV `cluster', which is a basic building block in MBQC \cite{zhou2003quantum,menicucci2006universal,proctor15measurement}.

For the qubit sub-case, it has been shown by Kashefi \emph{et al.}~\cite{kashefi2009twisted} that, up to local gates, the two interactions $E_{ar}$ and $\check{E}_{ar}$ are the only possible choices that allow for \emph{deterministic} universal quantum computation within the constraints of ADQC. The full range of possible interactions in higher dimensions has not been determined and is in general a difficult task. Adapting this $\check{E}_{ar}$ swap-based gate will provide the interaction for the model we present in the next section.

\section{Minimal control computation \label{Smincont} }
The ADQC model for general quantum variables we have presented in the previous section has a range of appealing features, including that it requires a minimal level of access to the computational register but may still implement universal quantum computation on it. However, it requires high-quality variable-basis measurements on the ancillas, and implementing these is intrinsically challenging in any quantum system. 

In this section we present an alternative model of ancilla-based quantum computation which may implement universal quantum computation using only
\begin{enumerate}
\item A fixed ancilla-register interaction gate, which may be applied to any ancilla-register pair.
\item Ancillas prepared in the computational basis.
\end{enumerate}
Because the model proposed here bypasses the need for on-line local controls of any kind on the register or on the ancillary QVs, it allows the entire set-up to be optimised for a high fidelity fixed ancilla-register interaction and long coherence times in the computational register. Obviously, these conditions are with the exception that some measurements must be performed at the end of the computation to read out the result (these may be performed on the register or on ancillas). This model is applicable to all types of QVs, although it is perhaps better suited to qubits and qudits than QCVs, as discussed later, and it includes as a special case a qubit-based model presented by some of these authors in Ref.~\cite{proctor2014minimal}.

As with the ADQC model, it is important to pick a suitable fixed interaction. Define a general two-QV diagonal gate, denoted $D(\phi)$ and parameterised by $\phi:\mathbb{S}_d^2 \to \mathbb{R}$, by
\begin{equation} 
\ket{q}_r|q'\rangle_s  \xrightarrow{D_{rs}(\phi) }  e^{i\phi(q,q')} \ket{q}_r|q'\rangle_s.
\end{equation}
The model we propose is based on a fixed ancilla-register interaction of the form
\begin{equation} 
\hat{E}_{ar}(u,\phi) :=  u_a \cdot  \textsc{swap} \cdot D_{ra}(\phi) ,
\end{equation}
with some unitary $u$ and some two-parameter function $\phi$, which are for now both left unspecified in the interests of flexibility. Note that this is a natural extension of the \textsc{swap}-based gate that may be used for ADQC, as given in Eq.~(\ref{check-S}), and because the interaction is based on \textsc{swap} it is applicable only when the ancillary and register QVs are of the same dimension.

\subsection{Implementing local and entangling gates}

It is straight-forward to confirm that the fixed interaction gate, when either the ancilla or the register QV is in a computational basis state, implements the mappings
\begin{align}
  \ket{\psi} \otimes \ket{q} &\xrightarrow{\hat{E}_{ar}(u,\phi) } \ket{q} \otimes u R(\phi(\cdot,q) ) \ket{\psi},  \label{swap1} \\ 
\ket{q} \otimes \ket{\psi} &\xrightarrow{\hat{E}_{ar}(u,\phi))}  R(\phi(q,\cdot) ) \ket{\psi} \otimes u\ket{q}, \label{swap2}
\end{align}
where $\phi(\cdot,q)$ and $\phi(q,\cdot)$ are the one-parameter phase-functions obtained from $\phi$ with the first and second variables fixed to $q$, respectively. Therefore, if either QV is in a computational basis state then the gate acts as a $\textsc{swap}$ along with local gates. Hence, an entangling gate may be implemented on a register QV pair using only three interactions and an ancilla prepared in any computational basis state. 

In particular, it is simple to confirm that
\begin{equation} 
 \ket{\psi}_{rs}  \otimes \ket{0} \xrightarrow{\hat{E}_{ar}\hat{E}_{as}\hat{E}_{ar}} W_{rs}(u,\phi) \ket{\psi}_{rs} \otimes u\ket{0} ,
\label{Eminuni22}
 \end{equation}
where $W_{rs}(u,\phi)$ is the two-QV gate
\begin{equation} 
W_{rs}(u,\phi) = R_r(\phi(0,\cdot)) \cdot \hat{E}_{rs}(u,\phi) \cdot u_rR_r(\phi(\cdot,0)).
\label{def-W}
\end{equation}
The $W_{rs}(u,\phi)$ gate is entangling except for special choices of $\phi$, specifically it is entangling if there is some $q,q' \in \mathbb{S}_d$ such that 
\begin{equation} \phi(q,q) +\phi(q',q')- \phi(q,q') - \phi(q',q) \hspace{0.1cm} \text{mod} \hspace{0.1cm} 2\pi \neq 0, \end{equation}
 which is generically true.  This entangling gate implementation method may be summarised by the circuit diagram
  
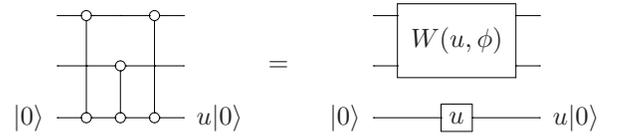
\begin{equation*}  
  \label{ent-cir-d}
  \Qcircuit @C=0.9em @R=1.0em {
    & \ctrlo{2} &  \qw &  \ctrlo{2} &  \qw &&&& && & &&\multigate{1}{W(u,\phi)} & \qw   \\
   & \qw & \ctrlo{1}  & \qw  &  \qw &&&&= &&&& &\ghost{W(u,\phi)} & \qw  \\
  \lstick{\ket{0}} & \controlo \qw  & \controlo \qw  & \controlo \qw &  \rstick{u\ket{0}} \qw  &&&&&&& &  \lstick{\ket{0}}& \gate{u} &  \rstick{u\ket{0}} \qw   
 }\vspace{0.3cm}
 \end{equation*}
where, as earlier, two quantum wires connected via a line and two `$\circ$' symbols is used to denote the fixed interaction gate (which is now $\hat{E}_{ar}(u,\phi)$, rather than $E_{ar}$). Note that the three gates used here to entangle a pair of register QVs via an ancilla is one more than needed in the ADQC model, but it is the minimum possible using unitary dynamics alone \cite{lamata2008sequential}.

A set of $|\mathbb{S}_d|$ different gates may be implemented on any register QV, with the gate chosen by specifying the preparation state of an ancilla and interacting it twice with the register QV. More specifically, from Eqs.~(\ref{swap1} -- \ref{swap2}) it follows that
\begin{equation} \ket{\psi} \otimes \ket{q} \xrightarrow{\hat{E}_{ar}\hat{E}_{ar} } s(q)\ket{\psi} \otimes u \ket{q} ,
\label{Eminuni21}
 \end{equation}
 where $s(q)= R(\phi(q,\cdot ) ) u R(\phi(\cdot,q) )$. This gate technique may be summarised in the circuit diagram
  
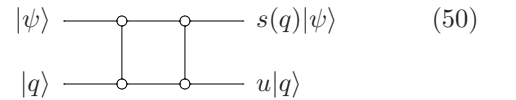
\begin{equation}
  \label{s-circ-}
  \Qcircuit @C=2.0em @R=2.0em {
\lstick{\ket{\psi}}& \ctrlo{1}& \ctrlo{1} &   \rstick{s(q) \ket{\psi}} \qw  \\
 \lstick{\ket{q}}  & \controlo \qw & \controlo \qw  & \rstick{u\ket{q} } \qw   \\
 }\vspace{0.3cm}
 \end{equation}
 Note that the price of using a swap-based interaction without the aid of measurements is that two gates are required to implement each $s(q)$ local unitary.

 \subsection{Universal gate sets}
The two gate methods proposed above allow the deterministic implementation of the gate set
   \begin{equation}
  \mathcal{G} = \{ W(u,\phi) ,s(q) \mid q \in \mathbb{S}_d \},
   \end{equation} 
   on the register QVs. If $W(u,\phi)$ is entangling, this gate set is sufficient for universal quantum computation if the single QV gates in the set are a universal set of single QV gates. This clearly is not the case for all choices of $u$ and $\phi$ (e.g., if $u$ is diagonal it cannot be universal, as then all of the $s(q)$ gate are diagonal), but for QVs that are qudits of any dimension choices of $u$ and $\phi$ can be found such that this set \emph{is} universal. A physically practical choice is given in Appendix~\ref{SC-gen}, and we conjecture that generic $u$ and $\phi$ are sufficient for universality. 
   
    Finally, we note that in some physical settings a universal set of high-fidelity local gates (i.e., single QV gates) may be available on the ancillas -- even if high-quality variable basis measurements are highly challenging. The \textsc{swap}-based model presented in this section may be optimised to this setting, allowing almost complete flexibility in the form the interaction may take to obtain universality. In particular, in this case any choices for the $\phi$ and $u$ parameters in $\hat{E}(u,\phi)$ such that the interaction is entangling are sufficient for universality, and ancillas need only be prepared in $\ket{0}$. This is because an entangling gate on the register may be implemented as above (see Eq.~(\ref{Eminuni22})) and any local gate may be implemented via applying a local gate \emph{to an ancilla} in between interactions of that ancilla with the register QV (i.e., adapting Eq.~(\ref{Eminuni21})). More specifically, to apply $v$ to the $r$\textsuperscript{th} register QV the gate $v' =R(-\phi(0,\cdot) )vR(-\phi(\cdot,0) )u^{\dagger}$ is applied to an ancilla QV prepared in $\ket{0}$ in between the application of two $\hat{E}_{ar}$ gates. That is,
   \begin{equation} 
\ket{\psi} \otimes \ket{0} \xrightarrow{\hat{E}_{ar}v'_a\hat{E}_{ar} } v \ket{\psi} \otimes u \ket{0} ,
 \end{equation}
 with this relation confirmed using Eqs.~(\ref{swap1} -- \ref{swap2}).
     This can be understood as an extension to general quantum variables of the qubit-based `ancilla-controlled quantum computation' model presented in \cite{proctor2013universal}. Note that if $u=\mathbb{I}$ then each gate has no overall effect on the ancilla that mediates it (it returns to the $\ket{0}$ state) and hence the ancillas may be reused to apply further gates to the register. However, using fresh or reinitialised ancillas prevents the propogation of correlated errors and is likely to be preferable in practice.

\section{Physical implementation}
In this penultimate section we briefly discuss the physical settings to which the ADQC and `minimal control' models might be particularly suited. The \textsc{cz} gate may be generated by the Hamiltonian $\hat{H}_1= \hat{x} \otimes \hat{x}$ applied for a time $t=2\pi/\hbar d$, and hence an $\hat{H}_1$ interaction between an ancilla and a register QV followed by fixed local $F$ and $F^{\dagger}$ gates (on the register and ancilla, respectively) implements the interaction $E_{ar}$. These local Fourier gates may be a fixed element in the experimental set-up, as they are applied after every interaction via $\hat{H}_1$, and they may be particularly simple in some cases. For example, with optical QCVs the Fourier gate and its inverse may be implemented with suitable length phase/time delays. 

In the context of qudits, it is more conventional to consider `spin' operators instead of $\hat{x}$ and $\hat{p}$. A qudit of dimension $d$ is a spin $s=(d-1)/2$ particle with a $z$-spin operator defined by $\hat{s}_z = \sum_{q \in \mathbb{S}_d} \frac{2q+1-d}{2}\ket{q}\bra{q}$. As $\hat{x}=\hat{s}_z +(d-1)\mathbb{I}/2$, letting $\hat{x} \to \hat{s}_z$ in $\hat{H}_1$ results in a Hamiltonian, $\hat{H}_2 = \hat{s}_z \otimes \hat{s}_z$, which still generates \textsc{cz} (up to local rotation gates). We now consider which physical systems might be particular suited to the ADQC and `minimal control' models, considering QCVs, qudits and hybrid QVs in turn.

Considering QCVs, the two ancilla-register interactions proposed herein for the ADQC model ($E_{ar}$ and $E_{ar}'$) are both Clifford (i.e., Gaussian) and hence, if the register and the ancillary QCVs are both realised in optics, either of these interaction can be composed from a fixed circuit of beam splitters and local Gaussian transformations \cite{braunstein2005quantum}. This is promising as many Gaussian transformations on optical QCVs are routine experimental techniques \cite{weedbrook2012gaussian}. However, it should be noted that some of these tranformations must be active optical elements as $E_{ar}$ and $E_{ar}'$ do not preserve total photon number. However, the QCV ADQC model is perhaps more advantageous (in comparison to, say, a direct implementation of MBQC) in the setting of atom-based QCVs: a computational register could consist of matter-based `quantum memory' QCVs interfaced via ancillary optical QCVs. Indeed, there have been some experiments along these lines -- optical QCVs have been both stored in \cite{jensen2011quantum}, and used to entangle \cite{gross2011atomic,krauter2013deterministic}, atomic-ensemble QCVs. 

Assuming the most physically relevant setting, whereby the ancillas are realised optically, homodyne measurements and photon-number-resolving detection (PND) of these ancillas is sufficient for universal QCV ADQC, as shown in Section~\ref{finite-s}. Encouragingly, Clifford gates driven by homodyne detection have already been demonstrated \cite{ukai2011demonstration,su2013gate} in the context of QCV MBQC, and there have been significant recent improvements in PNDs \cite{humphreys2015tomography,calkins2013high}, suggesting that QCV non-Clifford gates may be realisable soon. The final important resource required in this setting is highly squeezed input ancillas. The current experimental record is 12.7 dB of squeezing in optical states \cite{eberle2010quantum,mehmet2011squeezed}, but it seems likely that squeezing nearer 20 dB will be required for computations of indefinite length in QCV ADQC, as this is the known squeezing threshold for (qubit-encoded) fault-tolerant QCV MBQC \cite{menicucci2014fault}.

We now consider possible settings that might be suitable for realising the higher dimensional qudit ADQC model and the \textsc{swap}-based `minimal control' model of Section~\ref{Smincont} -- we are neglecting the qubit special case as that has been discussed elsewhere \cite{anders2010ancilla,proctor2014minimal,proctor2013universal,Proctorthesis2016,proctor2016hybrid,halil2014minimum} for these and closely related models. Impressive controls and high quality measurements of higher dimensional qudits have been realised in a range of physical systems, including superconducting \cite{neeley2009emulation} (up to $d=5$), atomic \cite{smith2013quantum,anderson2014accurate} ($d=16$) and photonic systems \cite{bent2015experimental,walborn2006quantum,lima2011experimental,rossi2009multipath,dada2011experimental}  (up to $d=12$), where in the optical case the qudit is encoded in the linear \cite{lima2011experimental,rossi2009multipath} or orbital angular momentum (OAM) \cite{bent2015experimental,dada2011experimental} of a single photon. Alternatively, a qudit may be encoded into the collective excitations of a qubit ensemble, with ensembles of this sort having been realised using, for example, caesium atoms \cite{christensen2013quantum} and nitrogen-vacancy centres in diamond \cite{zhu2011coherent}. Such ensembles have been investigated as a possible long-life quantum memory for qubits \cite{rabl2006hybrid,marcos2010coupling,lu2013quantum,petrosyan2009reversible} but they may also be used to store qudits (or QCVs as discussed above). This suggests that atomic-ensembles might be a suitable setting for a low-decoherence computational qudit register. Interestingly, OAM-encoded qubits have been stored in such atomic ensembles \cite{ding2015quantum}, and the technique of Ref.~\cite{ding2015quantum} can in principle be used to store OAM qudits for $d>2$. Hence, given that only experimental constraints limit the dimensionality of the qudits which may be encoded into OAM, and a range of high-quality measurements have been demonstrated on OAM-encoded qudits \cite{bent2015experimental}, OAM qudits may be particularly well-suited to mediating gates on a register of atomic-ensemble based qudits, with very high values of $d$ possible in principle.

Another possible encoding for qudits is into quantum harmonic oscillators, using the first $d$ energy eigenstates of a quantum harmonic oscillator as the computational basis of the qudit \cite{bartlett2002quantum}. In this setting, the qudit case of the $\hat{H}_1$ Hamiltonian may be implemented if two oscillators can be coupled via the Hamiltonian $\hat{H}_{3} = \hat{a}^{\dagger}\hat{a} \otimes \hat{b}^{\dagger}\hat{b}$, and hence a $\textsc{cz}$ gate may be generated with an appropriate evolution time. $\hat{H}_3$ is often called the cross-Kerr Hamiltonian, and it has been engineered using electromagnetically induced transparencies \cite{sun2008phase,yang2009enhanced}, optical fibres \cite{matsuda2009observation,li2005optical} and cavity QED systems \cite{mucke2010electromagnetically,zhu2010large}.  

Finally, we briefly consider the `hybrid' setting in which the ancillary and register QVs are of different types, noting that in this setting the ADQC model is only guaranteed to be universal in a stochastic sense (see Section~\ref{Ascz}). With a register of qubits interfaced via QCV ancillas, the Hamiltonian $\hat{H}_4= \sigma_z \otimes (\hat{a} + \hat{a}^{\dagger})$ may be used to generate hybrid \textsc{cz} gates, which are often called `controlled displacements' \cite{spiller2006quantum}, and high-quality interactions of this sort have been realised in superconducting systems \cite{wang2009coupling,xue2012fast,yoshihara2016superconducting}. Alternatively, the dispersive limit of the Jaynes-Cummings model \cite{jaynes1963comparison}, with a qudit encoding into the quantum harmonic oscillator, generates a hybrid qubit-qudit \textsc{cz} gate \cite{proctor2014quantum} and this regime of the Jaynes-Cummings model has been experimentally realised in \cite{schuster2007resolving,wallraff2004strong}.

\section{Conclusions}

We have presented a model for universal quantum computation in which only very limited access is required to a well-isolated computational register and the computation is driven via measurements of ancillas. Furthermore, this model has been formulated to be directly applicable to qubits, higher dimensional qudits and QCVs.  To be more specific, in this model universal quantum computation is implemented on a register using \emph{only} repeated applications of a single fixed two-body gate (which may be applied to any ancilla-register pair) and variable basis measurements of the ancillas which are prepared in a fixed initial state. This includes as the qubit special case the so-called \emph{ancilla-driven quantum computation} (ADQC) model \cite{anders2010ancilla,kashefi2009twisted}, and for this reason the same terminology has been used herein. Because measurement outcomes are fundamentally probabilistic, the measurements of the ancillas introduce random Pauli errors into the computation. Nonetheless, step-wise determinism is possible using classical feed-forward of measurement outcomes, in a similar fashion to measurement-based quantum computation (MBQC). 

We have shown that the parallelism inherent in the MBQC model is also available in the ADQC model we have proposed here, for all types of quantum variable -- i.e., for qudits of any dimension and QCVs. This includes the power to implement \emph{any} circuit of Clifford gates in essentially one quantum computational layer, which is not possible using unitary quantum gates alone  \cite{proctor15measurement}. Hence, the ADQC model is not only appealing from a practical perspective but it is also powerful for parallel quantum computation. The measurement bases that are sufficient for universal quantum computation have been discussed and in particular we showed that in the setting of QCVs, with the ancillas realised as optical states, homodyne detection and photon-number counting are sufficient for universality. This is promising, as homodyne detection is now a routine quantum optics technique \cite{ukai2011demonstration,su2013gate} and there have been many recent improvements in photon-number-resolving detectors \cite{humphreys2015tomography,calkins2013high}.

We then presented a \emph{globally unitary} ancilla-based model, which may be more relevant in settings in which high-quality measurements on ancillas are challenging or not possible. In this `minimal-control' model, universal quantum computation may be implemented using only a single fixed ancilla-register interaction and ancillas prepared in states from the computational basis. Hence, being unable to perform measurements to drive the computation has been compensated for with state preparation, using the natural symmetry between state preparation and projective measurement. 

The models presented herein allow for a computational register to be fully optimised for long coherence times and a single interaction with some ancillary systems, which may be physically distinct and chosen for their convenient properties (e.g., natural interactions with the register systems). Universality is then obtained via very limited manipulations of the ancillary systems. Hence, we have provided methods for realising universal quantum computation on a well-isolated register with a practical and simple scheme that is applicable to qubits, higher dimensional qudits and quantum continuous variables.

\section*{Acknowledgments}
We thank T. Nakano for the preliminary results on translating ADQC to the QCV regime. TJP was funded by a University of Leeds Research Scholarship. MG and NK acknowledge support from the Scottish Universities Physics Alliance (SUPA) and the International Max Planck Partnership (IMPP) with Scottish Universities. VK is funded by UK Engineering and Physical Sciences Council Fellowship EP/L022303/1.

\appendix

\section{\label{AAscz}}
In this appendix we briefly outline the minor changes that are required to the gate methods in the ADQC model when the fixed ancilla-register interaction is the gate
\begin{equation} 
\check{E}_{ar}:= F_a \cdot \textsc{swap} \cdot \textsc{cz},
\label{check-S2}
 \end{equation}
rather than $E_{ar}= F_rF_a^{\dagger} \textsc{cz}$, which was used throughout the main text. The two-QV gate implemented by sequential interactions of an ancilla with QVs $r$ and $s$ followed by an $\hat{x}$ measurement may easily be confirmed to be 
\begin{equation}
\frac{ \bra{m} \check{E}_{as}  \check{E}_{ar}  \ket{+_0}}{\|\bra{m} \check{E}_{as}  \check{E}_{ar}    \ket{+_0}\| }
=   X_s(-m) F_rF_s C^r_sX,
\end{equation}
where $m\in\mathbb{S}_d$ is the measurement outcome. Note that this implements a slightly different entangling gate on the register to when the interaction is $E_{ar}$ (see Eq.~(\ref{ADQCent}) and Eq.~(\ref{ADQCent2})) but it is still a Clifford gate.

The same set of single QV gates (i.e., any $FR(\vartheta)$ gate) can be implemented using this alternative interaction by measuring slightly different operators. Specifically, an interaction of an ancilla with a register QV followed by a measurement of $\hat{x}_{F R(\vartheta) F^{\dagger}}$ on the ancilla implements $FR(\vartheta)$ up to a Pauli error as
\begin{equation}
 \frac{ \bra{m} F_a R_a(\vartheta) F_a^{\dagger}  \check{E}_{ar}  \ket{+_0}}{\|\bra{m} F_a R_a(\vartheta) F_a^{\dagger}  \check{E}_{ar}    \ket{+_0}\| }
=  X_r(-m) F_rR_r(\vartheta).
\end{equation}
Although the gate set that may be implemented with this interaction is not identical to the one implemented with the $E_{ar}$ interaction (the entangling gate is different), the same techniques of classical-feedforward may be used to implement the computation deterministically. The only difference is a minor change in the exact form of the required classical side-processing.

\section{ \label{SC-gen}}

In this appendix it is shown that,  for any dimension of qudit, there are choices for the parameters $u$ and $\phi$ in the gate set
   \begin{equation}
  \mathcal{G} = \{ W(u,\phi) ,s_{u,\phi}(q) \mid q \in \mathbb{S}_d \},
   \end{equation} 
such that it is universal for quantum computation, where $W(u,\phi)$ is defined in Eq.~(\ref{def-W}) and $s(q)$ is given by $s(q)= R(\phi(q,\cdot ) ) u R(\phi(\cdot,q) )$). We conjecture that generic choices for the parameters $u$ and $\phi$ will be sufficient for universality for all dimensions of qudit, and it may be possible to confirm this using similar ideas to those used in \cite{lloyd1995almost}. However, here we provide a more specific choice for $u$ and $\phi$, for which we explicitly prove universality.

Let $u=F$ and take any $\phi$ such that $\phi(q,q')= 0$  for all $q,q' \in \mathbb{Z}(d)$ except when $q'=d-1$, in which case $\phi(q,d-1)=\theta_q$ with $\theta_{q}$ randomly (and independently) sampled from $\mathbb{R}$ for all non-zero $q \in \mathbb{Z}(d)$ and $\theta_0=0$. It is immediately clear that in this case $W(u,\phi)$ is entangling, and hence showing that the set of local gates $s(q)$ with $q \in \mathbb{Z}(d)$ can approximately generate (to arbitrary accuracy) any local gate is sufficient to prove universality. It is easily confirmed that $s(0)=F$. It is therefore also possible to implement the gates $s(q)s(0)^3=R(\phi(q,\cdot ))$ for $0<q<d-1$. Because $\phi(q,q')=\theta_q$ for $q'=d-1$, and $\phi(q,q')=0$ otherwise, then this gives a method for implementing a gate which applies no phase to all the basis states except the $\ket{d-1}$ basis state, for which it applies a `generic' phase (which is different for each $q$). Because these phases are generic, it is therefore possible to approximate any gate which applies only a phase to this last basis state to arbitrary accuracy. Now, $s(d-1)= R(\phi(d-1,\cdot ) ) F R(\phi(\cdot,d-1) )$, and $R(\phi(d-1,\cdot ) ) $ is a gate which applies only a phase to the last basis state. Because with $s(q)$ gates with $q=0,\dots,d-2$, the gate $R(-\phi(d-1,\cdot ) ) $ can be implemented to arbitrary accuracy and $s(0)^3=F^{\dagger}$, it is possible to obtain the gate $s'(d-1)=R(\phi(\cdot,d-1) )$ from the available set. Now, $\phi(\cdot,d-1)$ is a generic phase function (Note that, although here the $\phi(0,d-1)=0$, i.e., only the other $d-1$ values of $\phi(\cdot,d-1)$ are `generic', this is irrelevant as this may be considered to be fixing the global phase of the rotation gate) as implied by the conditions on $\phi$ given above, and as a rotation gate with a generic phase function in combination with the $F$ gate (obtained as $s(0)$) is a universal set of single-qudit gates \cite{Proctorthesis2016,proctor15measurement} this confirms the universality of the available gate set with an interaction gate of this form.

 The construction given above may seem rather contrived, however it represents a physically sensible gate -- a $D_{ra}(\phi)$ gate with $\phi$ as described above is a gate which implements phases on the register qudit only if the ancilla qudit is in the state $\ket{d-1}$. However, if this model were to be of further interest (outside the qubit-based setting, in which further appropriate choices for $u$ and $\phi$ can be found in \cite{Proctorthesis2016,proctor2014minimal}) it would be important to undertake a more thorough investigation of which parameter choices in the interaction are sufficient for universality. Finally, note that universality in the QCV model has not been investigated as it does not seem likely that this model will be of practical interest in this case. One reason for this is that Gaussian (i.e., Clifford) operations are generally much simpler to implement than non-Gaussian operations in the most promising QCV setting of optics (e.g., a Gaussian entangling gate can be achieved via a beam-splitter). Hence, in this setting it makes more sense to consider a Gaussian computer aided by some non-Gaussian operator used as sparingly as possible and this does not fit into the paradigm considered here, whereby a quantum computer is based entirely on a single gate which must be non-Gaussian to achieve universality.

\section*{References}
\bibliographystyle{apsrev}
\bibliography{Bib_Library}

\end{document}